\documentclass[%
 reprint,
 amsmath,amssymb,
prl,
aps,
showpacs,
]{revtex4-1}

\usepackage{graphicx}% Include figure files
\usepackage{dcolumn}% Align table columns on decimal point
\usepackage{bm}% bold math
\usepackage{amssymb}
\usepackage{amsfonts}
\usepackage{amsmath}
\usepackage[colorlinks,hyperindex]{hyperref}
\input{epsf}
\begin{document}

\title{Periodic Array of Bose-Einstein Condensates in a Magnetic Lattice}

\author{ S.~Jose$^{1}$, P.~Surendran$^{1}$, Y.~Wang$^{1}$, I.~Herrera$^{1}$, L.~Krzemien$^{2}$, S.~Whitlock$^{3}$, R.~McLean$^{1}$,
A.~Sidorov$^{1}$  and P.~Hannaford$^{1}$}
\email[]{phannaford@swin.edu.au}
\affiliation {\makebox[\textwidth]{$^1$Centre for Quantum and Optical Science, Swinburne University of Technology, Melbourne, Australia 3122.}
\makebox[\textwidth]{$^2$Jerzy Haber Institute of Catalysis and Surface Chemistry, Polish Academy of Sciences, 30-239 Krakow, Poland.}\\
 $^3$Physikalisches Institut, Universit$\ddot{a}$t Heidelberg, Im Neuenheimer Feld 226, 69120 Heidelberg, Germany}

\date{\today}

\begin{abstract}

 We report the realization of a periodic array of Bose-Einstein condensates of $^{ 87}$Rb \textit F=1 atoms trapped in a one-dimensional magnetic lattice close to the surface of an atom chip. A clear signature for the onset of BEC in the magnetic lattice is provided by \textit{in-situ} site-resolved radiofrequency spectra, which exhibit a pronounced bimodal distribution consisting of a narrow component characteristic of a BEC together with a broad thermal cloud component. Similar bimodal distributions are found for various sites across the magnetic lattice. The realization of a periodic array of BECs in a magnetic lattice represents a major step towards the implementation of magnetic lattices for  quantum simulation of many-body condensed matter phenomena in lattices of complex geometry and arbitrary period.
\end{abstract}
\maketitle
\begin{figure}[htbp]
	\begin{center}
		\includegraphics[width=0.5\textwidth]{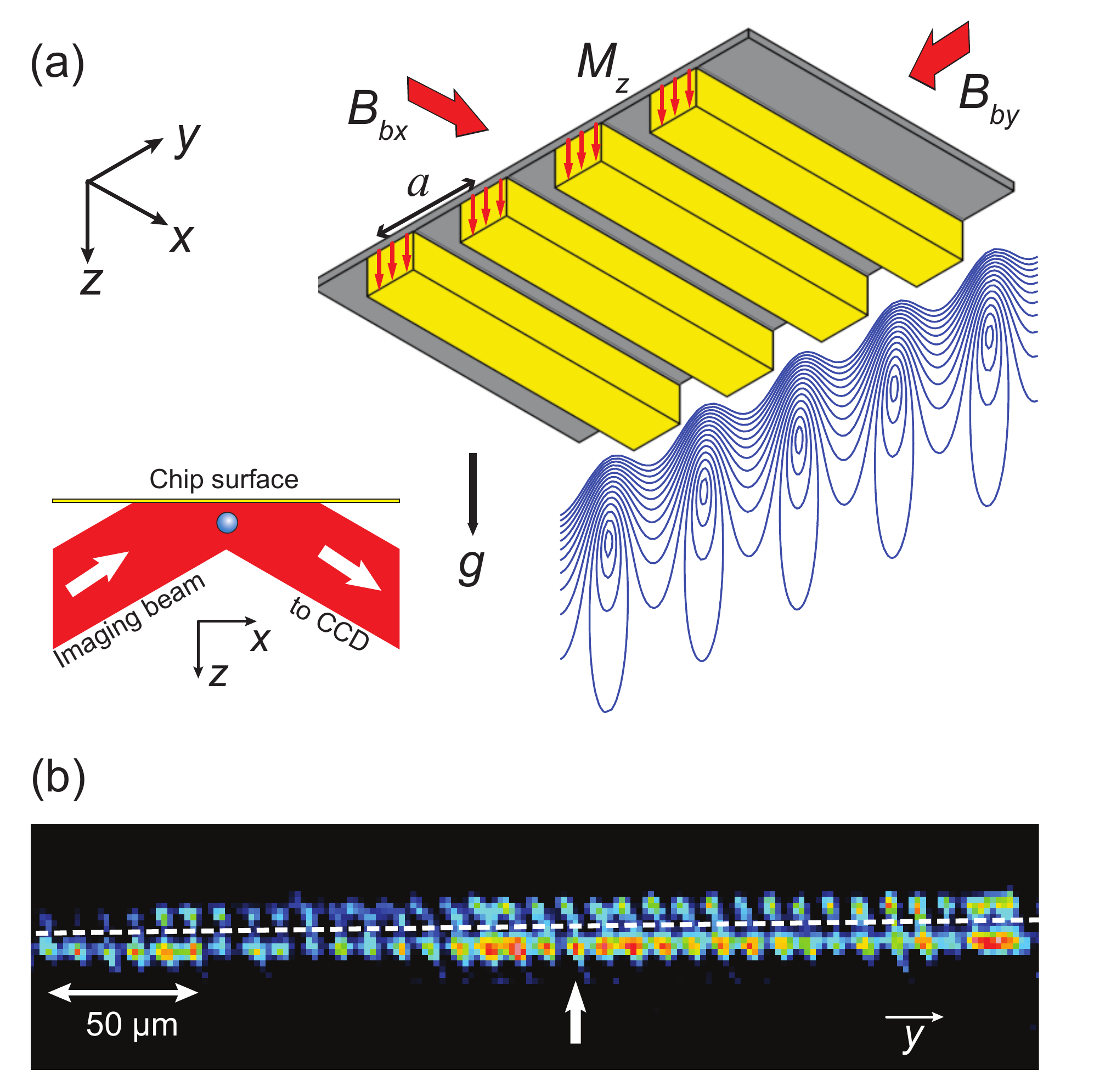} 
		\caption{(colour online)(a) Schematic of the magnetic microstructure used to create a periodic 1D lattice of magnetic microtraps with non-zero potential minima.~Contour~lines are equipotentials calculated for the parameters in this experiment with contour interval of  0.5 G. (b) Part of absorption image for an array of clouds of $^{87}$Rb $\vert \mathrm{\textit F=1,\textit m_F=-1}>$ atoms trapped in the 1D 10 $\mu$m-period magnetic lattice, after evaporative cooling to a trap depth $\delta f$=$(f_{f}$-$f_{0}$)=100 kHz, which cools the atoms below the critical temperature. The image was recorded by reflective absorption imaging along the axial \textit x-direction (see inset), which produces images both after reflection (bottom image) and prior to reflection (top image) of the imaging beam from the chip surface (dashed line). The effective pixel size is 2.0 $\mu$m. The vertical arrow indicates the lattice site (site 38) at which the RF spectra in Fig. \ref{fig:fig2} were recorded.}  
                      \vspace{-1.9em}                    
		\label{fig:fig1}
	\end{center}
\end{figure}

Optical lattices based on arrays of optical dipole traps are used extensively to trap periodic arrays of ultracold atoms and quantum degenerate gases in a broad range of applications. These range from simulations of condensed matter phenomena \cite{Bloch08} to studies of low-dimensional quantum gases \cite{Kinoshita06}, high precision atomic clocks \cite{Takamoto05} and registers for quantum information processing \cite{Calarco00,Monroe07}. A potentially powerful alternative approach involves magnetic lattices based on periodic arrays of magnetic microtraps created by permanent magnetic microstructures \cite{Ghanbari06,Gerritsma06, Gerritsma07,Boyd07,Singh08,Whitlock09, Schmied10,Abdelrahman10,Llorente10, Leung11,Leung13}, current-carrying wires \cite{Yin02,Grabowsk03,Gunnther05} or vortex arrays in superconducting films \cite{ Romero13}. Magnetic lattices based on patterned magnetic films may, in principle, be tailored to produce 2D (or 1D) arrays of atomic ensembles in arbitrary configurations \cite{Schmied10}. Periodicities may range from tens of micrometers, i.e., the interesting range for Rydberg-interacting quantum systems, such as Rydberg-dressed BECs \cite{Honer10} and Rydberg-mediated quantum gates \cite{Leung11,Saffman10}, down to below the optical wavelength where tunneling coupling strengths may exceed those possible with conventional optical lattices. Currently, there is also much interest in creating 2D periodic lattices of complex geometry, such as triangular, honeycomb, Kagome and super-lattices, in order to simulate condensed matter phenomena \cite{Lewenestein07}, including exotic quantum phases, such as graphene-like states \cite{Zhu07,Tarruell,Uehlinger13}, which are predicted to occur in lattices with non-cubic symmetry.\\
\indent Despite these prospects for magnetic lattices, little has been achieved to date, compared to optical lattices, in part due to the difficulty in controlling the resulting potentials, including magnetic homogeneity and efficient loading of the microtraps. Another serious challenge is to overcome the inelastic collision losses which can occur at high atom densities and which are accentuated when miniaturizing the traps. For example, previous experiments involving 2D arrays of magnetic microtraps with a period of about 25 $\mu$m \cite{Whitlock09} were limited by rapid three-body loss (decay rates $>$20 s$^{-1}$) which precluded the formation of Bose-Einstein condensates with observable condensate fractions.\\
\indent In this Letter we report clear signatures for the realization of a periodic array of Bose-Einstein condensates (BECs) of $^{ 87}$Rb atoms in a 1D 10 ${\mu}$m-period magnetic lattice. The signature for the onset of BEC is provided by \textit{in-situ} site-resolved radio-frequency (RF) spectroscopy \cite{Whitlock09,Whitlock07}. To minimize three-body losses in the magnetic lattice the atoms are prepared in the $|\mathrm{\textit F=1, \textit m_F=-1}>$ low-field seeking state which has a three-times smaller three-body recombination coefficient \cite{Burt97,Soeding99} and weaker magnetic confinement than the $\vert \mathrm{\textit F=2, \textit m_F=+2}>$ state on which previous work was based \cite{Whitlock09}. Additionally, we employ lattice traps with lower trap frequencies and hence lower peak atom densities. The realization of a periodic array of BECs in a magnetic lattice represents a major step towards creating periodic arrays of BECs in more complex lattice geometries with smaller lattice periods which are required for simulation of many-body condensed matter phenomena.\\
\indent Our 1D lattice of magnetic microtraps is created by superimposing the magnetic field from a grooved, perpendicularly magnetized magnetic film with uniform bias fields $B_{bx}$ and $B_{by}$ along the \textit x- and \textit y-directions (Fig.~\ref{fig:fig1}a) \cite{Ghanbari06,Singh08}.~The magnetic field components in the case of an infinite 1D lattice with no axial confinement along the \textit x-direction and for distances ${z>>a/2\pi}$ from the magnetic microstructure are [$B_{x}, B_{y},B_{z}$] $\approx$ [$B_{bx}; B_{0}$sin($ky$)e$^{-kz}$+$B_{by}$; \textit B$_{0}$cos($ky$)e$^{-kz}$]  \cite{Ghanbari06}, where $a$ is the lattice period, \textit k=2$\pi/a$, $B_0$=4$M_z(e^{kt}$--1), \textit t is the magnetic film thickness, and $M_z$ is the magnetization~(in Gauss).~A contour plot of calculated equipotentials, which are proportional to ${\vert\bf B}(x,y,z)\vert$, is shown in Fig.~\ref{fig:fig1}a. The strength and direction of the bias fields determine the potential minima, their distance from the magnetic microstructure, the trap frequencies, and the barrier heights \cite{Ghanbari06}. The potential minima need to be non-zero to prevent losses due to Majorana spin-flips.
\\
\indent Details of the 1D 10 $\mu$m-period magnetic microstructure and atom chip are described elsewhere \cite{Singh08,SJose13}. Briefly, the microstructure consists of a 10 mm$\times$10 mm magnetic film deposited on a microfabricated grooved silicon substrate on the atom chip, which is mounted upside-down in the UHV chamber. The magnetic film is a six-layer structure of 1.0 $\mu$m-thick perpendicularly magnetized Tb$_6$Gd$_{10}$Fe$_{80}$Co$_4$ (total thickness 1.3 $\mu$m), for which $4\pi M_z$$\sim$3 kG, $ H_\mathrm C$$\sim$6 kOe and $T_\mathrm{Curie}$$\sim$300$^o$C \cite{Wang05}. The magnetic microstructure is mounted 300 $\mu$m below a combined $U$- and $Z$-wire oriented with its central section perpendicular to the grooves. The bias fields and weak axial confinement are provided by the current-carrying \textit Z-wire \cite{SJose13}. For $I_z$=17 A and $B_{bx}$=51 G, the trap frequencies determined from numerical simulations are $\omega _x/2\pi$=259 Hz and $\omega_{y,z}/2\pi$=7.3 kHz, which is consistent with $\omega_y/2\pi$=7.5 kHz measured by parametric heating. The trap frequencies correspond to a geometric mean frequency  $\bar{\omega}/2\pi= (\omega_x\omega_y^2)^{1/3}/2\pi$=2.40 kHz and an aspect ratio of 30. In comparison to earlier experiments on a 2D magnetic lattice \cite{Whitlock09}, $\bar{\omega}$ is about four times smaller and, correspondingly, the expected three-body loss rates, which scale as $\bar{\omega}^6$ \cite{Whitlock14}, are about 10$^4$ times smaller. The calculated barrier heights are $\Delta B_y$=4~G (130 $\mu$K) and $\Delta B_z$$\sim$1 G (30 $\mu$K). 

Typically, 1$\times$10$^8$ $^{87}$Rb atoms are collected in a mirror MOT 1.2 mm below the chip before being transferred to a compressed $U$-wire MOT where they are polarization-gradient cooled and optically pumped to the $|\mathrm{\textit F=1, \textit m_F=-1}>$ state. The atoms are then transferred to a $Z$-wire trap 600 $\mu$m below the chip where they are evaporatively cooled. About 3$\times$10$^6$ atoms at 10-15 $\mu$K are then brought close to the magnetic lattice by ramping $I_z$ from 38 A down to 17 A in 100 ms with $B_{bx}$= 51 G. Under these conditions the \textit Z-wire trap merges smoothly with the magnetic lattice microtraps located 8 $\mu$m below the chip, allowing $\sim$1$\times$10$^6$ atoms to be loaded into the magnetic lattice \cite{SJose13}.

The lifetime of the atoms in the magnetic lattice microtraps is $\sim$12 s, which is sufficient for the atoms to be evaporatively cooled (for 1.5 s) by ramping an RF field from 7.0 MHz down to a final evaporation frequency $f_f$. A trap depth of $\delta f$=$(f_f-f_0$)=100 kHz (where $f_0$ is the trap bottom) leaves $\sim$5$\times$10$^4$ atoms trapped in $\sim$100 lattice sites, or $<$N$>$$\sim$500 atoms per site, in the central region of the lattice.

Figure \ref{fig:fig1}b shows part of an \textit{in-situ} absorption image for a periodic array of clouds of $^{87}$Rb $|\mathrm{\textit F=1, \textit m_F=-1}>$ atoms trapped in multiple sites of the 1D 10 $\mu$m-period magnetic lattice after evaporative cooling to a trap depth $\delta f$=100 kHz, which cools the atoms below the critical temperature. The image was recorded using reflective absorption imaging  \cite{Armijo10} along the long axis of the elongated atom clouds. The imaging beam is focused by a cylindrical lens into a light sheet and sent at a slight angle ($\sim$2$^o$) to the reflecting gold surface of the chip, resulting in images both after and prior to reflection. Detection of the two images provides a measure of the distance of the trapped atoms from the chip surface  (8 $\mu$m). The effective pixel size is 2.0 $\mu$m, corresponding to 5.0 pixels per lattice period. The measured resolution from the width of the images of individual sites is 4 $\mu$m. 

The clouds of atoms in Fig.~\ref{fig:fig1}b are resolved in their individual lattice sites, which allows us to perform site-resolved measurements. The variation in site-to-site transmitted intensity ($\sim\pm$30$\%$) across the lattice  is due mainly to imperfections in the gold mirror on the chip surface and to non-uniform loading of the lattice and non-uniformity of the imaging light. In addition, some of the variation is due to inhomogeneity in the magnetic lattice. The positions of the individual clouds of atoms reveal that the lattice period is constant to within 1$\%$ across the lattice.
\begin{figure}[htbp]
	\begin{center}
		\includegraphics[width=0.5\textwidth]{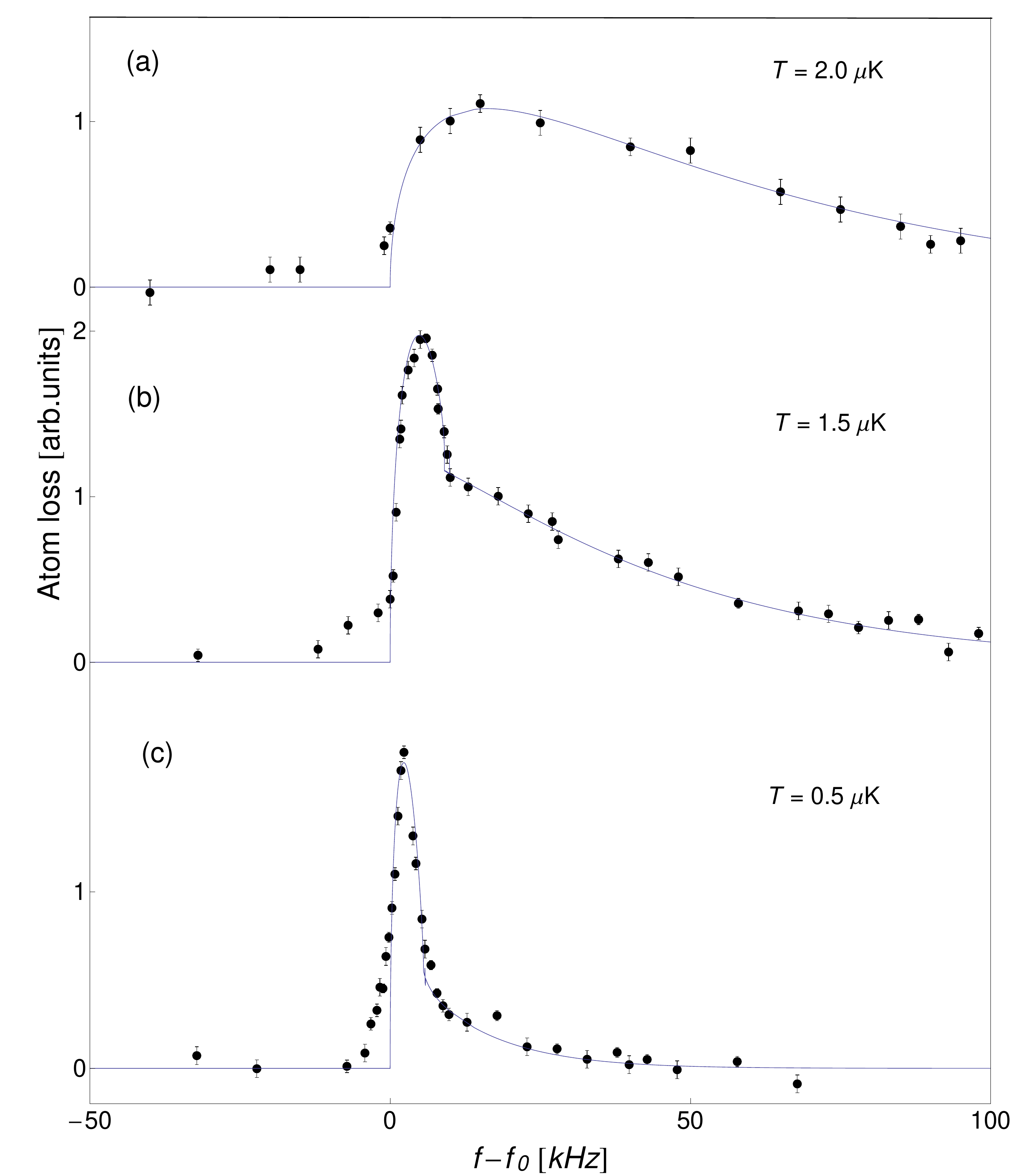} 
		\caption{RF spectra for $^{87}$Rb $\vert \mathrm{\textit F=1,\textit m_F=-1>}$ atoms after evaporative cooling in the 1D magnetic lattice to trap depths $\delta f=(f_f-f_0)$ of (a) 600 kHz, (b) 400 kHz and (c) 100 kHz in lattice site 38. The solid lines represent fits to the data based on a self-consistent mean-field model for a BEC plus thermal cloud as described in the text.}
                       \vspace{-1.5em}                
		\label{fig:fig2}
	\end{center}
\end{figure}

Figure \ref{fig:fig2} presents \textit{in-situ} RF spectra of atom loss over a range of RF frequencies \textit f taken at a single lattice site (site 38, vertical arrow in Fig.~\ref{fig:fig1}b) after the atoms have been evaporatively cooled to trap depths $\delta f$=600, 400 and 100 kHz. The power of the RF pulse was reduced to one-tenth of that used for evaporative cooling to minimize power broadening, and the RF pulse duration was 40 ms. The RF spectra evolve from (a) a broad truncated Boltzmann type frequency distribution for $\delta f$=600 kHz, characteristic of a thermal cloud, through to (c) a narrow Thomas-Fermi type frequency distribution for $\delta f$=100 kHz, characteristic of a BEC. For (b), $\delta f$=400 kHz, the RF spectrum exhibits a pronounced bimodal distribution consisting of BEC plus thermal cloud components \cite{Fernholz08}. The critical temperature for quantum degeneracy for \textit N=500 atoms per site ($\delta f$=100~kHz) and $\bar{\omega} /2\pi$=2.40 kHz is $T_\mathrm c$=0.9 $\mu$K. The time decay of the BECs in the magnetic lattice corresponds to a half-life of $\sim$0.5~s.

To fit the data in Fig. \ref{fig:fig2}, we use a self-consistent mean-field model which takes account of atom-atom interactions in both the BEC and thermal cloud and the mutual interaction between them \cite{Whitlock09}, but neglects the kinetic energy of the condensate fraction via the Thomas-Fermi approximation and effects of gravity sag in the tightly confining magnetic traps. For the density distribution of the condensates we use \cite{Gerbier04}
\begin{equation} \label{eq:1}
n_\mathrm{c}(r)=\mathrm{Max}{\{(\frac{1}{g})[\mu-V(r)-2gn_\mathrm{th}(r)];0}\}
\end{equation}
where~$n_\mathrm{th}(r)$=Li$_{3/2}[exp(-\vert \mu\mathord{-}V_\mathrm{eff}(r)\vert) /k_\mathrm{B}T]/\lambda_\mathrm{dB}^3$~is~the density~distribution~of~the~thermal~cloud, $V_\mathrm{eff}(r)$=$V(r)$+$2g[n_\mathrm{th}(r)$+$n_\mathrm{c}(r)]$,~and~${V(r)}$=1/2\textit M($\omega _x^2x^2+\omega _y^2y^2+\omega _z^2z^2)$ is the confining harmonic potential. Li$_{3/2}$[z] is the polylogarithmic function with base 3/2, $\lambda$$_{dB}$ is the thermal de Broglie wavelength, \textit g=4$\pi\hbar^2a_s/M$ is the mean-field coupling constant, $M$ is the atom mass, and $a_s$ is the s-wave scattering length.~ Equation~\ref{eq:1} is solved iteratively to obtain $n_\mathrm{c}(r)$ for a given temperature \textit T and chemical potential $\mu$, where the elongated cloud with trap frequencies $\omega_x$ and $\omega_y$ can be replaced by a spherical cloud with mean trap frequency $\bar{\omega}$=$(\omega_x\omega_y^2)^{1/3}$. The RF frequency distribution is then obtained from the resonance condition $hf$=$\mu_B|g_FB|$ for $\Delta m_F$=$\pm$1 spin-flip transitions and by determining the number of atoms $\Delta$$N(f)$ in a spherical shell 4$\pi\textit r^2\Delta\textit r$ of constant frequency $f$ that are removed by a RF knife of width $\Delta$$f$. For the case of a pure BEC, the number of trapped atoms removed by the RF knife at frequency \textit f is $\Delta N(f)\propto \mathrm{Max} \{f^{1/2}(\mu/(\vert m_\mathrm F\vert h)\mathord{-}f)\Delta f; 0\}$ with a base-width $\mu/(\vert m_\mathrm F\vert h)$.

The solid lines in Fig. \ref{fig:fig2} represent fits to the RF spectra using the above model. For a given data set, $\mu$ and \textit T are varied until the calculated RF spectrum provides a reasonable fit to the experimental spectrum with the constraint that the total  atom number $N_{\mu ,T}$ derived from $\mu$ and $ T$ is consistent with the atom number \textit N$_\mathrm{abs}$ derived from the absorption image and normalized to $<$$N_{\mu,T}$$>$ for various lattice sites. For the trap depth $\delta f$=400 kHz (Fig.~\ref{fig:fig2}b), the data fit well to a narrow BEC component plus a broad thermal cloud components with chemical potential $\mu/h$=19 kHz, corresponding to N$\sim$8700 atoms, and temperature \textit T=1.5~$\mu$K. The slight rounding of the leading edge of the spectrum is attributed to residual power broadening. For the trap depth $\delta f$=100~kHz (Fig.~\ref{fig:fig2}c), the fit yields $\mu /h$=7.5 kHz, corresponding to \textit N$\sim$380 atoms, and \textit T=0.50~$\mu$K. The condensate fraction deduced from the fit is 50 (5)$\%$, or 54 (5)$\%$ if we include the contribution from the rounded leading edge in Fig.~\ref{fig:fig2}c.

\begin{table}[b]
\vspace{-1.5em}
\caption{\label{tab:table1}
Trap bottom ($f_0$), chemical potential ($\mu/h$), atom temperature (\textit T), condensate fraction ($\%$) and atom number (\textit N$_\mathrm{abs})$, derived from absorption images and normalised to $<$\textit N$_{\mu, T}$$>,$ for various sites across the lattice. The trap depth $\delta f$ is 100 kHz.
}
\begin{ruledtabular}
    \begin{tabular}{llllll}
    Site & $f_0$ (MHz)         & $\mu/h$ (kHz)        & \textit T ($\mu$K)         & BEC (\%)    & \textit N$_\mathrm{abs}$(atoms) \\
    22       & 4.9318 (5) & 7.0 (1.0)   & 0.55  (5) & 37 (5) & 240 (100)   \\
    35       & 4.9320 (5) & 7.5 (0.5) & 0.50 (5)      & 50 (5) & 510 (50)    \\
    36       & 4.9322 (5)     & 7.5 (0.5)       & 0.50 (5)      & 50 (5)     & 430 (50)          \\
    37       & 4.9323 (5)     & 7.7 (0.5)       & 0.58 (5)      & 39 (5)     & 430 (50)         \\
    38       & 4.9322 (5)     & 7.5 (0.5)       & 0.50 (5)      & 50 (5)     & 460 (50)          \\
    39       & 4.9322 (5)     & 7.3 (0.5)      & 0.52 (5)      & 45 (5)     & 490 (50)          \\
    40       & 4.9324 (5)     & 7.5 (0.5)       & 0.55 (5)      & 42 (5)     & 480 (50)         \\
    41       & 4.9324 (5)     & 7.0 (0.5)       & 0.55 (5)      & 37 (5)     & 370 (50)         \\
    42       & 4.9324 (5)     & 7.0 (0.5)       & 0.55 (5)      & 37 (5)     & 400 (50)         \\
    43       & 4.9320 (5)     & 8.0 (0.5)       & 0.60 (5)      & 39 (5)     & 320 (50)         \\
    44       & 4.9325 (5)     & 7.5 (0.5)       & 0.53 (5)      & 45 (5)     & 430 (50)         \\
    67       & 4.9320 (5)     & 7.8 (0.5)       & 0.60 (5)      & 38 (5)     & 440 (50)         \\
    70       & 4.9319 (5)     & 7.2 (0.5)       & 0.56 (5)      & 38 (5)     & 390 (50)          \\
    71       & 4.9320 (5)    & 7.8 (0.5)       & 0.55 (5)      & 45 (5)     & 460 (50)         \\
    85       & 4.9320 (5) & 8.7 (1.0) & 0.60 (5)  & 44 (5) & 450 (100)   \\
     \end{tabular}
\end{ruledtabular}
\end{table}
Our absorption imaging scheme allows RF spectra to be recorded simultaneously for all of the $\sim$100 populated sites across the magnetic lattice, with a total acquisition time of about one hour. The spectrum for each of the analyzed lattice sites exhibits a strong BEC component, similar to Fig.~\ref{fig:fig2}c. Table \ref{tab:table1} summarizes the results for a sample of 15 sites which are representative of the central  region (sites 23-85) of the lattice, including a string of 10 adjacent sites (35-44), and sites near the ends of the central region. The site-to-site variations in $f_0$, $\mu$, \textit T and condensate fraction indicate that the sites are remarkably uniform across the lattice. In particular, the trap bottoms, which could be accurately determined from the intercepts of the fitted  RF spectra with the ($f-f_0$) axis, show site-to-site variations of only $\pm$ 0.4 kHz corresponding to $\pm$0.6 mG in 7.0 G. This degree of uniformity across the lattice is not reflected in the absorption image in Fig.~\ref{fig:fig1}b or in the atom numbers $N_\mathrm{abs}$ derived from the absorption image  which show significant variations in transmitted intensity as discussed above. 

As an independent check for the onset of Bose-Einstein condensation in the magnetic lattice we measure the atom loss rate due to three-body recombination. For a BEC, the density-density fluctuations are suppressed relative to a thermal cloud, due to a factor of six smaller three-particle correlation function $g^{(3)}$ ~\cite{Burt97}. The three-body recombination loss rate is given by $dN(t)/dt$=$-K_3\int n^3({\textit{\textbf r}},t)d^3r$, where $K_3$ is the loss rate constant. Assuming that during the atom loss the BEC maintains a Thomas-Fermi profile, the atom number decay becomes $N_\mathrm c(t)$=$N_\mathrm0(t)(1+\mathrm {\beta} t)^{-5/4}$ where $\beta$=$(32/105)K_3 n^2_\mathrm{c,p}(0)$ and $n_\mathrm{c,p}(0)$ is the initial peak density of the condensate. The three-body recombination rate is measured by monitoring the atom loss for hold times out to 500 ms where the condensate still persists. A power-law
fit to the atom decay curve yields an exponent of~$\mathord{-}1.21(4)$, which is consistent with the BEC value of $\mathord{-}5/4$ but significantly different from the thermal cloud value of~$\mathord{-}1/2$. For an initial peak density $n_\mathrm{c,p}(0)$=$8\times$$10^{14}~\mathrm{cm}^{-3}$, which was determined from analysis of the RF spectra, the three-body decay constant is found to be $K_3$=$5.5~(1.0)\times$$10^{-30}~\mathrm {cm}^6\mathrm s^{-1}$, which is consistent with previous measurements for $^{87}$Rb $|\mathrm{\textit F=1, \textit m_F=-1>}$ atoms~\cite{Burt97,Laburthe04}, but much smaller than the value $K_3$=$4.2~(1.8)$$\times10^{-29}$ $ \mathrm{cm}^6\mathrm s^{-1}$ measured for a thermal cloud~\cite{Burt97}.

\indent Arrays of BECs in a 10 $\mu$m-period magnetic lattice are promising for the implementation of Rydberg-interacting quantum systems by exploiting the long-range dipolar interaction between atoms excited to Rydberg states. The size of each BEC in the array is well within the typical Rydberg blockade radius, so that each BEC could potentially be used as a collective qubit. The interaction driven level shift between two n$\approx$80s Rydberg excitations at a distance of 10 $\mu$m is still several MHz, which far exceeds the Rydberg state decay rate \cite{Saffman10}. At the same time, each trap is sufficiently far from the chip surface to minimize unwanted surface effects  \cite{Leung11}. It should also be possible to create spatially separated Rydberg-dressed BECs \cite{Honer10} or degenerate Fermi gases, in which Rydberg states are weakly admixed to the atoms, resulting in strong long-range and anisotropic interactions. This might enable the realization of \textquoteleft coupled quantum gases\textquoteright~where atoms in spatially separated traps may strongly influence one another (e.g., \cite{Lakomy12}).

\indent Another application will be to engineer simple graphene-like model systems with tunable parameters, for example, using magnetic lattices with hexagonal symmetry loaded from a BEC or degenerate Fermi gas. In periodic lattices the tunneling rates scale with period \textit a and barrier height $V_0$ as $J$ $\propto a^{-1/2}V_0^{ 3/4} exp[-CV_0^{1/2}a$] (where \textit C=(32\textit M)$^{1/2}/h$) \cite{Bloch08}. For a square optical lattice with \textit a=0.64 $\mu$m, the tunneling rate for $^{87}$Rb atoms is estimated to be \textit J$\sim$20 Hz for  $V_0$$\sim$$12E_r$ (where $ E_r$ is the recoil energy) \cite{Bakr09} or \textit J$\sim$330 Hz for $V_0$$\sim 6$$E_r$. Thus, in order to have significant tunneling rates in a magnetic lattice, sub-micron periods are required. Due to the tighter confinement, atomic states with low inelastic collision rates should be chosen (for example, fermionic atoms such as $^{40}$K) or the number of atoms per site should be limited to less than three, which would normally be the case for a 2D lattice with sub-micron period. Additionally, the magnetic potentials need to be smooth and homogeneous with constant periodicity ($<$1-2\% \cite{Romero13}) and uniform trap bottoms. To produce high-quality magnetic potentials with sub-micron periods we propose to use nano-fabricated multi-atomic layer Co/Pd (or Pt) films (with~$\sim$6 nm grain size) presently under development for high density data storage.\\
\indent In conclusion, we have realized a periodic array of Bose-Einstein condensates in a 10 $\mu$m-period 1D magnetic lattice. A clear signature for the onset of BEC was provided by \textit{in-situ} site-resolved RF spectra which show a pronounced bimodal distribution. Similar bimodal distributions were found for various sites across the magnetic lattice. This result represents a major advance towards the implementation of magnetic lattices to create periodic arrays of BECs for quantum simulation of many-body condensed matter phenomena in lattices of complex geometry and arbitrary period .\\
\indent We thank Mandip Singh, Brenton Hall and Chris Vale for fruitful discussions.~We acknowledge funding from an ARC Discovery Project grant (DP130101160).
\\

% {\bf Acknowledgements:}

\end{document}